\documentclass[journal]{IEEEtran}

\ifCLASSINFOpdf
\usepackage[pdftex]{graphicx}
\usepackage{color}

\usepackage{amsmath}
\usepackage{amsfonts} 
\usepackage{amsthm}

\newtheorem{theorem}{Theorem}
\newtheorem{lemma}{Lemma}

\usepackage[shortcuts,acronym]{glossaries}

\makeglossaries 

\newacronym{ai}{AI}{Artificial Intelligence}
\newacronym{bs}{BS}{Base Station}
\newacronym{fifo}{FIFO}{First In First Out}
\newacronym{isac}{ISAC}{Integrated Sensing and Communication}
\newacronym{tmt}{TMT}{target monitoring terminal}
\newacronym{twi}{TWI}{Temporal Window of Integration}
\newacronym{xr}{XR}{eXtended Reality}

\newtheorem{defn}{Definition}[section]

\begin{document}

\title{Time, Simultaneity, and Causality in
Wireless Networks with Sensing and Communications}

\author{Petar Popovski,~\IEEEmembership{Fellow,~IEEE}
\thanks{P. Popovski is with the Department of Electronic Systems, Aalborg University, Denmark.}
}

\maketitle

\begin{abstract}
Wireless systems beyond 5G evolve towards embracing both sensing and communication, resulting in increased convergence of the digital and the physical world. The existence of fused digital-physical realms raises critical questions regarding temporal ordering, causality, and the synchronization of events. This paper addresses the temporal challenges arising from the fact that the wireless infrastructure becomes an entity with multisensory perception. With the growing reliance on real-time interactions and applications such as digital twins, extended reality, and the metaverse, the need for accurate timestamping and temporal forensics becomes crucial. The paper introduces a model that incorporates Temporal Windows of Integration (TWI) to emulate human multisensory perception and discusses the implications for setting timing constraints in real-time applications and enabling temporal forensics. The analysis explores trade-offs, probabilities, and bounds for simultaneity and causality violation in the context of wireless systems evolving towards perceptive networks. This work underscores the significance of timestamping in the evolving wireless landscape, provide insights into system-level implications, and points out new research avenues for systems that combine sensing and communications.
\end{abstract}

\IEEEpeerreviewmaketitle

\section{Introduction}

\IEEEPARstart{T}{ime} discrepancy and manipulation has been a continuous inspiration to science fiction authors~\cite{wells2011time,dick2002minority}. The physical time has its order and arrow~\cite{rovelli2019order,reichenbach1956direction}, such that its manipulation is still in the domain of science fiction. Nevertheless, the perception of time in digital interconnected systems is dependent on the way information is processed and transported. Our reality becomes increasingly a fusion of physical and digital worlds, which brings in the central question of perception of time and ordering of events. 


With the advent of 5G wireless systems, the domain of digital communication was expanded to embrace real-time interaction among humans and machines, bringing into focus latency and other timing-related measures as some of the main design objectives~\cite{parvez2018survey,yates2021age}. The low latency constraints from 5G are set to evolve towards more general timing requirements in 6G \cite{popovski2022perspective}, reflecting the way information is processed and used.
There is a growing number of applications that rely on real-time interaction over digital communication links, such as digital twins~\cite{alkhateeb2023real}, \ac{xr}~\cite{akyildiz2022wireless} or the metaverse~\cite{ball2022metaverse}. The evolution vector of wireless systems beyond 5G points towards increased fusion of the digital and the physical world.
An early consensus reached about 6G has been the integration of communication with two other modalities for acquisition of information, sensing and localization~\cite{liu2022integrated}. This gives rise to the concept of \emph{perceptive mobile networks}~\cite{xie2023collaborative}, in which the mobile network infrastructure is used to map the physical into the digital world via sensing. 

The physical-digital convergence brings in the central questions of \emph{chronology}, or temporal ordering, and \emph{causality} among the intertwined physical and digital events. Specifically, the devices that use digital links, sensors, as well as  \acf{isac}~\cite{liu2022integrated} can be treated as entities with \emph{multisensory perception}. Digital links also contribute to the multisensory perception by carrying sensory data from the mobile devices or from dedicated, passive target monitoring terminals~\cite{xie2023collaborative}. In the 3GPP terminology~\cite{3gpp.22.837}, sensing using radio resources is termed 3GPP-sensing, while carrying sensory data through digital links is referred to as transparent sensing. Similar to the multisensory perception for humans~\cite{vroomen2010perception}, it is critical to have the proper temporal ordering and definition of simultaneity for the events coming from different sensor and links. 

\begin{figure}[t!]
 \centering
 \includegraphics[width=7.5cm]{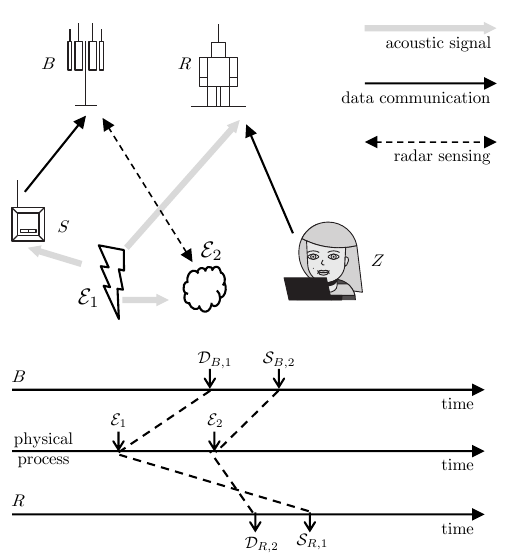}
 \caption{An example of chronological discrepancy. Acoustic event ${\cal{E}}_1$ causes the event ${\cal{E}}_2$ at a later time. $B$ receives ${\cal{E}}_1$ as ${\cal{D}}_{B,1}$, communicated by an acoustic sensor (microphone) $S$, and later on receives ${\cal{E}_2}$ as ${\cal{S}}_{B,2}$ by using its radar sensing. $R$ receives at first ${\cal{E}}_2$ as ${\cal{D}}_{R,2}$, an image sent by $Z$, and after that receives ${\cal{E}}_1$ as ${\cal{S}}_{R,1}$ via its own acoustic sensor.}
 \label{fig:ExampleScenario}
\end{figure}

Fig.~\ref{fig:ExampleScenario} provides a motivation for the ideas discussed in this paper. The acoustic event ${\cal{E}}_1$ causes the event ${\cal{E}}_2$, which can be a change in person's position or a change of the object's shape. The sensor $S$ has a microphone that registers ${\cal{E}}_1$ and transmits the data of this event to the \ac{bs} $B$. This is an instance of transparent sensing in 3GPP; throughout the paper we will refer to it as \emph{digital input}, in contrast to \emph{sensory input} obtained directly from a sensor. Hence, $B$ detects ${\cal{E}}_1$ as a reception of digital data ${\cal{D}}_{B,1}$. Furthermore, $B$ has a radar sensing capability and detects ${\cal{E}}_2$ as a sensing event ${\cal{S}}_{B,2}$. Here $B$ perceives the correct chronological and causal order of ${\cal{E}}_1$ and ${\cal{E}}_2$. The robot $R$ has an acoustic sensor and detects the event ${\cal{E}}_1$. Assuming that $R$ is at much larger distance to ${\cal{E}}_1$ as compared to $S$, it will detect ${\cal{E}}_1$ as a sensing event ${\cal{S}}_{R,1}$ with a significant delay. $Z$ sends an image of ${\cal{E}}_2$ to $R$, which corresponds to transparent sensing and is received by $R$ as an event ${\cal{D}}_{R,2}$. Thus, the chronological/causal order observed by $R$ is different from the one observed by $S$, that is, there is a \emph{chronological discrepancy} between $S$ and $R$. Considering the original causal order of ${\cal{E}}_1$ and ${\cal{E}}_2$, the robot $R$ experiences \emph{causality violation}. These occurrences can have significant impact in combined physical-digital worlds. For instance, 
if $Z$ is a malicious agent, she may preemptively send a synthetic, fake image to $R$ in order to affect the chronology/causality of the events. This may have dire consequences both in real-time, but also post-hoc, if forensics needs to be done after, say, traffic accidents that involve autonomous vehicles.

\begin{figure}[t!]
 \centering
 \includegraphics{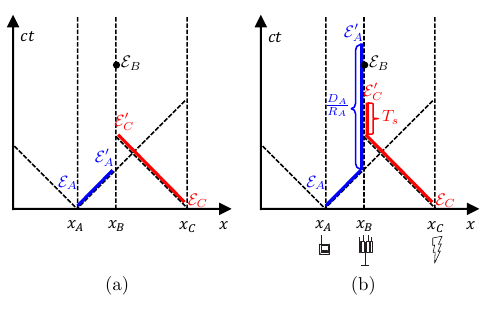}
 \caption{Illustration of light cones and the impact of information processing. The event ${\cal E}_i$ for $i=A, B, C$ occurs at the location $x_i$. ${\cal E}_A$ is received as ${\cal E}^{\prime}_{A}$ via digital link, while ${\cal E}_C$ is detected as ${\cal E}^{\prime}_{C}$ as a sensing event. The Base Station (BS) is placed at $x_B$. (a) Reception with instantaneous processing. The chronology observed by the BS is ${\cal E}^{\prime}_A \rightarrow {\cal E}^{\prime}_C \rightarrow {\cal E}_B$. (b) Reception of ${\cal E}_A$ with data size $D_A$ at a rate $R_A$ and detection of ${\cal E}_C$ with a sensor detection time $T_s$. The chronology observed by the BS is ${\cal E}^{\prime}_C \rightarrow {\cal E}_B \rightarrow {\cal E}^{\prime}_A$.}
 \label{fig:ConeDataProcessing}
\end{figure}

In physics, causal relations are related to the chronology of events, such that if the event ${\cal E}_1$ causes event ${\cal E}_2$, then it must be that ${\cal E}_1$ happened before ${\cal E}_2$, that is, at an earlier time. Conversely, if ${\cal E}_1$ happened before ${\cal E}_2$, then ${\cal E}_1$ does not necessarily cause ${\cal E}_2$. Thus, a temporal ordering of events only tells which causality relations are possible. However, two events at different spatial points may not be causally related in any way, such that one cannot determine which event happened earlier or whether the events have occurred simultaneously. This is one of the takeaways of the Special Theory of Relativity~\cite{einstein1905}, which is built upon the key assumption that the speed of light sets an upper bound on the speed by which information propagates through the spacetime and, in that way, it defines the possible causal connection between various events. Nevertheless, the Special Theory of Relativity does not consider the aspect of the finite rate of information transmission and processing, which results in additional time components that affect the perceived chronology of events. Fig.~\ref{fig:ConeDataProcessing} shows this impact using the light cones defined in Special Theory of Relativity. If the time required for information processing is ignored, then the chronology of the events observed at location $x_B$ is ${\cal E}^{\prime}_A \rightarrow {\cal E}^{\prime}_C \rightarrow {\cal E}_B$. However, if the data transmission rate from $A$ is low, the event ${\cal E}^{\prime}_A$ arrives later and the  chronology of the events at $x_B$ is changed to ${\cal E}^{\prime}_C \rightarrow {\cal E}_B \rightarrow {\cal E}^{\prime}_A$.

Inspired by relativistic time, Lamport brought the notion of \emph{logical clock} in networked computers and distributed systems~\cite{LamportClocks}. Logical clocks are used to attain temporal ordering of events in a distributed system, where different processes interact through communication links. The logical clock defines a consistent happened-before relation in the distributed system based on timestamps/counters, which can be used to define causal relations and histories. Looking at Fig~\ref{fig:ConeDataProcessing}(b), happened-before relations are defined by the \ac{bs} at $x_B$ based on the order of detection of the digital/sensing events. One may argue that we can easily go around the problem of variable data rate from $A$, as $A$ can be perfectly synchronized to the \ac{bs} and timestamp the transmitted data. However, the sensing event cannot be timestamped by the sender (nature) and it can only be timestamped after it enters into the digital systems. There are other tricks one can try to arrive to the ``true'' chronology, but there remains the principle of relativity that the chronology depends on the observer. Regarding causality, determining it cannot be always based on the times of arrival, as the example on Fig~\ref{fig:ExampleScenario} illustrates and one may need additional causal inference to determine the causal order of events. 

In this paper we consider the problem of temporal ordering and timestamping of events in a wireless infrastructure that involves communication and sensing. This infrastructure contains entry points for the sensing the physical into the digital world and, vice versa, for actuating commands from the digital into the physical realm. Hence, the future Base Stations and Access Points will have a \emph{timestamping functionality} that determines the chronology of the events, simultaneity, and causality. This functionality is used for two generic purposes: 
\begin{itemize}
    \item Facilitate real-time applications by setting timing constraints that are dependent on the multi-sensory contributions rather than fixed and predefined latency constraints. This is done in analogy with human multisensory perception, in which the \ac{twi} are used to determine which events occur simultaneously and thus determine the pace of reality at which humans reason and operate.
    \item Enable temporal forensics, not necessarily applied in real-time, to determine the chronology and causality of events. This may increase in importance as more synthetic data gets generated through \ac{ai}; indeed, in a world where manipulated data can impact the order of the events, with large societal and economic consequences, the wireless infrastructure will have the responsibility to, as consistently as possible, timestamp the physical reality.  
\end{itemize}
The specific contributions of this work are the following. \emph{(i)} Bring attention to the important problem of timestamping as wireless networks evolve towards being perceptive, by including sensing and communication. \emph{(ii)} Define a simple model that enables us to introduce the TWI and the notions of simultaneity/causality violation. \emph{(iii)}
Carry an elementary analysis for the case of two events to highlight the main tradeoffs in determining simultaneity and causality. This is followed by analysis for a sequence of $N$ events and derivation of probabilities and their bounds for simultaneity/causality violation. \emph{(iv)} System-level implications in terms of determining the latency requirements based on the TWIs and multisensory setup, as well as the resource allocation in ISAC-based systems. This is supplemented with new avenues for future research.

\subsection{Related Work}

\ac{isac} has been seen as one of the hallmarks of the 6G wireless systems~\cite{liu2022integrated}. Having different information acquisition modalities, as well as gathering transparent sensing data~\cite{3gpp.22.837} through digital transmissions, the \acp{bs} and the edge/mobile infrastructure evolves into an entity with multisensory perception. This has been the motivation for perceptive mobile networks from~\cite{xie2022perceptive}, where another layer of passive \acp{tmt} are added to the conventional cellular networks. this concept is further extended in~\cite{xie2023collaborative} where, among the other issues, the problem of clock synchronization is discussed, which is related to the present work.  

The related concepts from the literature on multisensory perception can be found in~\cite{spence2003multisensory, vroomen2010perception}, where the theme is inter-sensory sychnrony, which corresponds to the concept of simultaneity discussed in this paper.  

Another related body of work comes from the literature on logical clocks in distributed systems, applied to cyber-physical systems. The work~\cite{shrivastava2016time} considers timing in cyber-physical systems, the problem of synchronization and outlines the importance of chronology, simultaneity, and causality in these systems. 
A recent related work~\cite{lee2023consistency} shows how availability, a real-time property of a cyber-physical system, and consistency, a logical property, relate to clock synchronization and latencies introduced by networks and computation. They generalize the notion of consistency to include physical state rather than just variables in software.

\subsection{Paper Organization}

The introduction is followed by a System Model, which introduces the models for sensing and digital inputs, 
the timestamping function and the notions of simultaneity/causality violation. Section~\ref{sec:TwoInputs} analyzes the simultaneity and causality when only two inputs are considered, sensing or digital. This is generalized to $N$ inputs in Section~\ref{sec:Ninputs}. Section~\ref{sec:NumIllustration} provides a numerical illustration of the analysis presented before. This is followed by system-level perspective in Section~\ref{sec:SystemLevel} and discussion of timestamping, simultaneity and causality in practical wireless systems and Base Stations with \ac{isac}. Section~\ref{sec:discussion} discusses the possible generalization of the assumptions from this work and outlines avenues for future work. The last section concludes the paper and this is followed by appendices that contain proofs and supplementary analysis.

\begin{table} 
\caption{Table of Symbols}
\label{tableofsymbols}
\centering
\begin{tabular}{c|l}
\hline
\bfseries Symbols & \bfseries Description \\
\hline
${\cal E}$ & generic event, of any type \\
${\cal P}$ & physical event \\
${\cal S}$ & sensing event \\
${\cal D}$ & digital event \\
${\cal A}$ & actuation event \\
$\tau_a$ & action time \\
$T_s$ & sensor integration window \\
$D_s$ & data size produced by a sensor \\
$T_{p,s}$ & time between a physical event and its sensing event \\
$\tau_s$ & propagation time for a sensing event \\
$\phi_s$ & time offset for a sensing event \\
$T_{AB}$ & transmission time for digital data from $A$ to $B$ \\
$\tau_d$ & propagation time for digital data \\
$D_d$ & data size of a digital event \\
$W$ & size of the temporal window of integration \\
$s(t)$ & timestamping function \\
\hline
\end{tabular}
\end{table}

\section{System Model}

\subsection{Processes and Events}

Three event types are considered: physical, sensing, and digital.
A \emph{physical event} ${\cal P}$ happens in nature and, in the context of this work, where a physical event is any occurrence that can result in a detection by a sensor. A \emph{sensing event} ${\cal S}$ has a physical cause and happens at a digital system when a physical event/parameters are turned into digital data. Finally, a \emph{digital event} ${\cal D}$ has a digital cause and happens at a digital/computing system when a digitally encoded data is decoded and interpreted. The encoded data that causes a digital event may be, for example, sent over a communication link or generated locally at the computer. We use ${\cal E}$ as a generic notation of an event that can be of any type. A digital system, such as a mobile device or a wireless infrastructure, perceives the physical world through sensing and digital inputs. We note that there can be a fourth type of event, actuation event ${\cal A}$ that happens in nature by turning a digital data into a physical event/action; however, this is not discussed in details, as the focus of this work is on multi-input perception within the digital systems.  

The fundamental difference between a sensing and a digital event is that the sensing events are not labelled, while the digital events can be labelled. To illustrate, when a data is received over a communication link, it can be labelled by the ID of the sender and a timestamp when it was sent, for example, according to the local clock of the sender. However, a sensing event cannot be labelled, as nature does not add labels; the ID and the timing of the sensing event are subject to a process of inference and decision. For instance, an acoustic event does not carry a label about who or when created that sound; this can be inferred, with a certain confidence, through the data acquired by sensing that event.
It should be noted that the possibility for labeling and timestamping of the digital causes of the digital events is the basic premise for defining Lamport's logical clocks. 

A receiver, such as a \ac{bs} can have multiple inputs, where each input is either a sensory or a digital communication link. The events that come from the same input are received in a \ac{fifo} manner, such that the chronology of the events within a single input is preserved. 

We define \emph{action time} $\tau_a$ as the time required for an event at one location to cause an event at another location. For instance, in case of a lightning, $\tau_a$ is the time of light propagation plus the time for a sensor to detect the lighting. The action time for a digital event at device $A$ to cause a digital event at a device $B$ is the time needed to prepare the data packet at $A$, transmit it to $B$, and having $B$ decode the packet. 

\subsubsection{Physical Process} 

A physical process consists of a sequence of physical events. If not stated otherwise, throughout the paper it will be assumed that the events of a given physical process are causally related, such that a physical event affects causally all subsequent physical events from the same physical process.  

\subsubsection{Sensing Process}

A sensing process takes place at a sensor, defined here as a subsystem that converts physical parameters into a digital data. The sensing process consists of a sequence of sensing events, denoted by ${\cal S}_i$. A sensing event occurs after a sensor integrates the observation of the physical parameters for a duration equal or greater than the \emph{sensor integration window}, denoted by $T_s$. At the end of the 
window the sensor produces a data of size $D_s$ bits. This is a simplified model of how sensors work, as it represents the sensor as an entity that produces data after a certain integration time. This model is inspired by the \emph{dwell time} in radar systems~\cite{richards2010principles}, which is the time required by the radar to detect the target reliably. Our model idealizes the radar model by assuming a perfect target detection after a dwell time of $T_s$. The impact of the uncertainty and imperfection in the sensing process is discussed in~Section~\ref{sec:discussion}. 

\begin{figure}[t!]
 \centering
 \includegraphics[width=8.3cm]{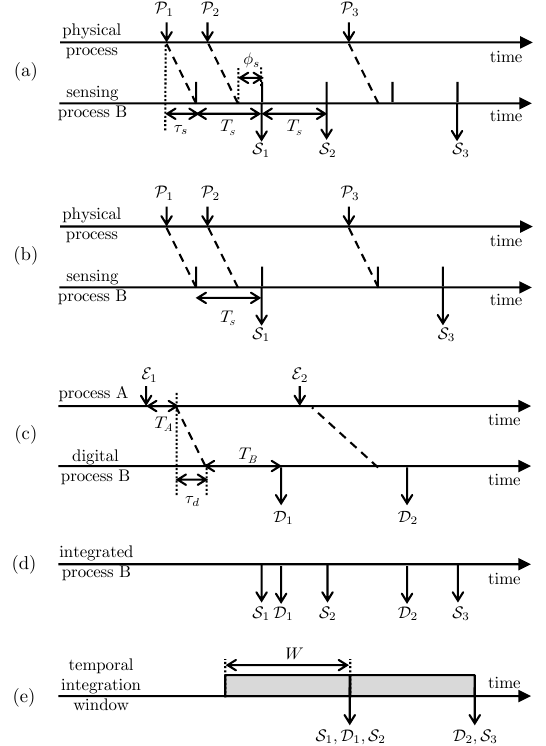}
 \caption{Illustration of temporal ordering in various processes. (a) Synchronous sensor. (b) Asynchronous sensor. (c) Communication between digital processes. (d) Integrated process of sensing and communication. (e) Temporal window of integration (TWI) applied to the integrated process from (d).}
 \label{fig:TemporalOrdering}
\end{figure}

Fig.~\ref{fig:TemporalOrdering}(a) depicts a 
\emph{synchronous sensor}, in which the sensor integration windows are consecutively ordered and the start of a window is not related to a physical event. This is a special case, as it assumes that the sensor integration window $T_s$ is equal to the periodicity of synchronous sensing and there is no idle/deaf time for the sensor. In principle, the periodicity window can be larger than $T_s$, such that the sensor has some idle time; this is further discussed in Section~\ref{sec:ISACtimestamping}. Assuming no idle time, the total time between the occurrence of the physical event and the detection by the sensor is:
\begin{equation}\label{eq:TotalTimeSensorEvent}
    T_{p,s}=\tau_s+\phi_s+T_s
\end{equation}
where $\tau_s$ is the time required to start the detection of the sensory event, not controllable by the sensor. In the simplest case it is equal to the \emph{propagation time} and this is how we will refer to it. In a more general case, it is the time until the event is detectable with the sensor available; for instance, an event becomes detectable by radar sensing upon causing motion or change of a position. In (\ref{eq:TotalTimeSensorEvent}), the time offset $\phi_s \in [0,T_s)$ between the instance at which the effect of the physical event becomes detectable by the sensor and the start of the integration window. We use $\phi_s$ to model the fact that the sensor is not event-driven, such that the detection needs to wait for new sensing window to start.  It therefore takes longer time for the sensor to detect ${\cal P}_2$ as ${\cal S}_2$ as compared to detecting ${\cal P}_1$ as ${\cal S}_1$. If the offset $\phi_s$ is uniformly randomly distributed in $[0,T_s)$, then $T_{p,s}$ is uniformly distributed in the support set $[\tau_s+T_s,\tau_s+2T_s)$. 

As illustrated in Fig.~\ref{fig:TemporalOrdering}(a), the sensor observes the physical process at a maximal rate of $R_s=\frac{D_s}{T_s}$ [bps] (bits per second); this happens when every event starts to be detected at the start of an integration window. This model does not capture analog sensors, such as human senses, but one can still define a data processing rate; see ``Bandwidth of senses'' in~\cite{norretranders1999user}. Note that, if a physical event causes a sensing event, then the action time is $\tau_a=T_{p,s}$.

The model can also be applied to an \emph{asynchronous sensor}, where the integration window is triggered by the arrival of a physical event. An asynchronous sensor needs to have a reliable mechanism that can trigger the start of a sensor integration window. As a result, we can model the asynchronous sensor by setting $\phi_s=0$ and thus there is a deterministic $T_{p,s}=\tau_s+T_s$. This is depicted on Fig.~\ref{fig:TemporalOrdering}(b) for the same physical events from Fig.~\ref{fig:TemporalOrdering}(a). 
Note that ${\cal S}_3$ is detected earlier as compared to the synchronous sensor on Fig.~\ref{fig:TemporalOrdering}(a). However, if the same interface is used to trigger a new integration window and perform detection during an integration window, then the arrival of new event cannot trigger a new window while another window is running; this is why the event ${\cal S}_2$ on Fig.~\ref{fig:TemporalOrdering}(b) is not detected at all. Such operation is captured by the simplified sensor model; in practice, the arrival of new events during an integration window would interfere with the detection of the original event.

The model can be used for both active and passive sensors. Specifically, the radar capability considered in ISAC is an active sensing, where a radio wave is transmitted and its reflections are processed during the integration window. Nevertheless, the model does not include the possible deaf time, during which the beam is not pointed to the direction of the event; see Section~\ref{sec:SystemLevel}.\ref{sec:ISACtimestamping}. 


\subsubsection{Digital Process}

A digital process consists of a sequence of events at a receiver of a digital communication link, as depicted on Fig.~\ref{fig:TemporalOrdering}(c). A \emph{digital event}, denoted by ${\cal D}_i$, at the receiver $B$ happens when a data packet, received over a digital input/link, is decoded. For simplicity, at the receiver $B$ we do not consider digital events that are locally generated by $B$'s computing system. At the sending device $A$ we denote the process simply as process $A$ and the events ${\cal E}_1, {\cal E}_2$ that initiate a transmission can be of two types: (1) physical event that is sensed by $A$ and transmitted digitally, i.e., a transparent sensing data. (2) digital event at $A$, created by decoding a command from an operating system, AI module, human input to a digital interface, or even a data receive through another digital link\footnote{In this case, if $A$ retransmits the same data received previously, we have a multihop transmission.}. {As mentioned before, unlike the sensing events at $B$, the digital events sent by $A$ are labelled; for instance, the sender $A$ can label the sensing data with the timestamp when the data was sensed.}  

The  transmission time between the event at the sender $A$ that initiated the digital transmission and the digital event at the receiver $B$ is:
\begin{equation}\label{eq:TotalTimeDigitalEvent}
    T_{AB}=T_A+\tau_d+T_B
\end{equation}
where $T_A$ is the time used by the sender $A$ until the data is encoded and transmission started, $\tau_d$ is the propagation time of the digital data, and 
$T_B$ is the time used by the receiver $B$ for decoding.  
The time $T_A$ depends on the type of event that triggers the transmission. If $A$ transmits sensing information to $B$, then $T_A$ consists of the time needed for $A'$s sensor to detect the event and the time for processing the data before the transmission.
Denoting the processing time after sensing by $T_{\mathrm{proc}}$, we have\footnote{It is interesting to relate this analysis to the popular measure of Age of Information (AoI)~\cite{kaul2012real,kosta2017age}. Assuming that the sensing data gets timestamped after sensing, AoI can be expressed as $T_{\mathrm{AoI}}=T_{\mathrm{proc}}+\tau_d+T_B$. Note that, for correct interpretation of the AoI, the clocks of $A$ and $B$ needs to be synchronized, otherwise $B$ cannot correctly interpret the timestamp added by $A$.}:
\[
T_A=\tau_s+\phi_{s}+T_s+T_{\mathrm{proc}}
\]
Differently from this, if $A$ transmits data that is synthetically generated by an AI algorithm, then ${\cal E}_i$ is a digital event, with zero propagation time, and $T_A=T_{\mathrm{proc}}$ is the processing time used for generating that data with the AI algorithm. 

For the digital sender/receiver we will not deal with the detailed operations and only assume that the digital event outputs a data of size $D_d$. Since we consider communication from devices to the wireless infrastructure, the propagation time $\tau_d$ corresponds to the light propagation time. In the more general case of multihop transmission, $\tau_d$ can be random due to jitter. For simplicity, we will assume that $D_d$ corresponds to a single packet sent over that link;  generalizations of this setup are given in Section~\ref{sec:discussion}.

The component $T_A$ is partially controlled by $A$, while $T_B$ is partially controlled by both $A$ and $B$. To see this assume that the encoding at $A$ and decoding at $B$ are instantaneous. Then  $T_A=0$ while $T_B=\frac{D_d}{R}$ where $R$ is the transmission rate over the digital link between $A$ and $B$ and this rate is subject to capacity constraints and codebook agreement by $A$ and $B$. For the subsequent analysis it will be more convenient to assume the equivalent setup in which the entire digital processing time $T_d=T_A+T_B$ takes place at the receiver $B$, such that an event at the digital process of the sender is immediately followed by the propagation time. 

For the subsequent analysis, we can model $T_{AB}$ as a random variable with probability density $p_{AB}(t)$ and non-negative support set $[T_{\min}, T_{\max}]$, such that $p_{AB}(t)=0$ outside of this set.


\subsection{Temporal Windows for Timestamping}

We now turn to the way in which the receiving device perceives the temporal ordering of the events. As illustrated on Fig.~\ref{fig:TemporalOrdering}(a) and (b), the sensing events coming from the same physical process or digital events based on the transmissions from the same sender preserve the causality of events. A more interesting situation occurs when there are multiple sensing and digital processes at a single device, as in this case it becomes questionable whether the causality from the physical world can be preserved for the events in the digital world. Fig.~\ref{fig:TemporalOrdering}(c)  depicts the \emph{integrated process} observed at $B$, where it is assumed that $B$ harbors both the sensor from Fig.~\ref{fig:TemporalOrdering}(a) and the digital receiver from Fig.~\ref{fig:TemporalOrdering}(b). It can be seen that ${\cal S}_1$ occurs before ${\cal D}_1$, while from the temporal relation between ${\cal E}_1$ and ${\cal P}_1$ it could be the case that ${\cal E}_1$ causes ${\cal P}_1$. 

This situation can be remedied by taking inspiration from multisensory perception in humans~\cite{vroomen2010perception}. For instance, looking at my hand and listening while I am knocking on a wood gives an impression of simultaneity, that is, the audio, visual, and tactile inputs from the knocking are perceived as simultaneous. This is because the brain uses a \emph{\acf{twi}} \cite{vroomen2010perception,spence2003multisensory} and all the sensing events that have occurred during a single window are treated as simultaneous. Similar to this, one can define a temporal integration window $W$ for a digital device that has multiple physical and digital inputs. This is depicted on Fig.~\ref{fig:TemporalOrdering}(d), where all events that happened during a temporal integration window are considered to have happened simultaneously. For the example, ${\cal S}_1, {\cal D}_1, {\cal S}_2$ happen simultaneously and are followed by the simultaneous occurrence of ${\cal D}_2, {\cal S}_3$ at the end of the next temporal integration window. 

We use the concept of \ac{twi} to  define a timestamping function $s: \mathbb{R} \mapsto \mathbb{Z}$ that assigns an integer timestamp to each received event:
\begin{equation}
    s(t)=\left\lceil \frac{t}{W} \right\rceil
\end{equation}
Here $W$ is the temporal integration window.
From a system perspective, a temporal integration window resides on top of the modules that receive the sensing and communication data and delivers timestamped events to the higher layers. Thus, if two events arriving at times $t_1$ and $t_2$ are received within the same temporal integration window, they have $s(t_1)=s(t_2)$, such that they are simultaneous from application perspective.



\subsection{Logical Clocks and Causality Violation}

In \cite{LamportClocks}, Lamport defined the \emph{happened-before} relation between two events. Based on this, there are three possible relations between two events ${\cal E}_1$ and ${\cal E}_2$ in a distributed system~\cite{baquero2016logical}: (1) ${\cal E}_1$  happened before and may have caused ${\cal E}_2$, denoted as ${\cal E}_1 \rightarrow {\cal E}_2$; (2) ${\cal E}_2$  happened before and may have caused ${\cal E}_1$, denoted as ${\cal E}_2 \rightarrow {\cal E}_1$; (3) ${\cal E}_1$ and ${\cal E}_2$ occurred concurrently, that is, it cannot be determined which one happened before, denoted as ${\cal E}_1 || {\cal E}_2$. With the introduction of the temporal integration window, we need to expand the set of possible relations to include the \emph{happens-simultaneously-with} relation; this is equivalent to the notion of logical simultaneity from~\cite{lee2023consistency}, where two events are simultaneous in a device if their logical timestamps are identical\footnote{The work~\cite{lee2023consistency} uses the term tag as more general that the timestamp, which enables consideration of a superdense time system; this is not in the scope for the current paper and we use only \emph{timestamp}.}. Thus, a receiver can perceive events in two ways:
\begin{itemize}
    \item If the receiver registers the event ${\cal E}_i$ before ${\cal E}_j$, then ${\cal E}_i$ \emph{happened before} ${\cal E}_j$ and this is denoted by ${\cal E}_i \rightarrow {\cal E}_j$.
    \item If the receiver registers the event ${\cal E}_i$ simultaneously with ${\cal E}_j$, then the event ${\cal E}_i$ \emph{happens simultaneously with} ${\cal E}_j$ and this is denoted by ${\cal E}_i \leftrightarrow {\cal E}_i$.
\end{itemize}
We can thus describe the events on  Fig.~\ref{fig:TemporalOrdering}(d) as ${\cal S}_1 \rightarrow {\cal D}_1 \rightarrow {\cal S}_2 \rightarrow {\cal D}_2 \rightarrow {\cal S}_3$, while for Fig.~\ref{fig:TemporalOrdering}(e) we have ${\cal S}_1 \leftrightarrow {\cal D}_1 \leftrightarrow {\cal S}_2 \rightarrow {\cal D}_2 \leftrightarrow {\cal S}_3$. 
Based on these definitions, two events registered at the same receiver cannot be concurrent. 

Let ${\cal E}_1$ and ${\cal E}_2$ denote two events and assume that ${\cal E}_1$ causes ${\cal E}_2$. Here ${\cal E}_i$ can be a physical, sensing, or digital event. In the following we assume that, whenever ${\cal E}_1$ happened before ${\cal E}_2$ then ${\cal E}_1$ caused ${\cal E}_2$, such that we can write ${\cal E}_1 \rightarrow {\cal E}_2$ to denote causality in the physical world. Let ${\cal E}^{\prime}_{A,i}$ denote the perceived event at the receiving device $A$ that corresponds to ${\cal E}_i$, where ${\cal E}^{\prime}_{A,i}$ can be either a sensing or a digital event. We can then define the following:   
\begin{defn} \label{def21}
Causality violation ${\cal V}_c$ for device $A$ occurs when two events are causally related as ${\cal E}_1 \rightarrow {\cal E}_2$, while their respective per events at $A$ are related as  ${\cal E}^{\prime}_{A,2} \rightarrow {\cal E}^{\prime}_{A,1}$. Simultaneity violation ${\cal V}_s$ for device $A$ occurs when two events ${\cal E}_1, {\cal E}_2$ should be perceived as simultaneous, while their respective received events at $A$ are related as  ${\cal E}^{\prime}_{A,2} \rightarrow {\cal E}^{\prime}_{A,1}$ or  ${\cal E}^{\prime}_{A,1} \rightarrow {\cal E}^{\prime}_{A,2}$.
\end{defn}

When the receiver is clear from the context, we will omit the subscript $A$. According to this definition, there is no causality violation if ${\cal E}_1 \rightarrow {\cal E}_2$, while ${\cal E}^{\prime}_{A,1} \leftrightarrow {\cal E}^{\prime}_{A,2}$. As an example, ${\cal E}_1$ can be an explosion ${\cal P}_1$, ${\cal E}^{\prime}_{A,1}={\cal S}_1$ can be a sensing event at $A$ that detects the explosion acoustically.
Another device, $B$ is much closer to the explosion and detects the sensing event ${\cal E}^{\prime}_{B,1}$ earlier than $A$ and sends a digital image of the explosion to $A$. Then the event ${\cal E}_2$ can be the reaction of a device $B$ to that explosion and ${\cal E}^{\prime}_{A,2}={\cal D}_2$ the received digital image of that reaction. A case where causality violation could be a problem is the one in which $B$ is a malicious device that, upon detecting the explosion, sends a fake image. If this fake image is generated and sent sufficiently quickly, a causality violation can occur and thus a doubt about the cause of the explosion. 

The following remark is in order. It is clear that causality/simultaneity violations are possible when there is at least one sensing event. When all the events are digital, they can be labeled at the senders $\{A_i\}$ and this can avoid causality violation, as the times of the initiating events will be adjusted. However, if the transmission at a device $A_i$ is initiated by a sensing event, then the true timing of the physical event that caused it still remains elusive. If, on the other hand, transmissions are initiated by digital events that are precisely timestamped at the senders, then causality violation can be avoided. However, this puts high demands on the required synchronization that may not be feasible for all devices. For instance, synchronization can be markedly improved by  using the Precision Time Protocol (PTP)~\cite{eidson2006measurement}; however, as stated in~\cite{lohstroh2023logical}, this \emph{``affects utility, but not correctness''}. If the clocks of the sender and the receiver are not synchronized perfectly and there is an offset of $\Delta$, then the probability  of causality/simultaneity violation depends on this $\Delta$. We will analyze the violation probabilities by not assuming any timestamping at the sending device; this can be further generalized, as discussed in Section~\ref{sec:discussion}. 

The introduction of \ac{twi} mitigates the causality violation, as the events are treated as simultaneous. If the events are timestamped as simultaneous, the causality is determined by a certain form of inference over the set of detected events, that is, the time component is insufficient to determine causality. Interestingly, the human brain also uses compensatory and inference mechanisms to determine sequence of events~\cite{vroomen2010perception}. The \ac{twi} introduces the following tradeoff. The larger the window $W$ is, the lower the chances of causality violation. However, larger $W$ decreases the temporal resolution of events, leading to a lower \emph{event throughput}. In fact, we can identify an \emph{event throughput loss} for a certain input as  $\frac{W}{T_0}$, where $T_0$ is the minimal detectable time between two events in a given sensory or digital input. For instance, the minimal detectable time for a sensor is $T_0=T_s$, the sensory integration time, while the minimal detectable time for a digital link is the minimal duration of a transmission $T_{\min}$.

\section{Simultaneity and Causality Violation with Two Inputs}
\label{sec:TwoInputs}

In this section we analyze several selected elementary cases. The analysis is done for simultaneity and causality violation according to the Definition~\ref{def21}. Note that for each elementary case both simultaneity and causality violation can be analyzed; however, for simplicity, only one type of analysis is shown per elementary case. Specifically, we analyze the probability of occurrence of these events under a certain stochastic model. A higher probability of simultaneity/causality violation is an indicator for a higher probability of incorrect operation of the overall system. Based on the presented analysis, the reader can apply similar analysis to more general cases. 

\subsection{Simultaneity with Two Sensory Inputs}

\begin{figure}[t!]
 \centering
 \includegraphics[width=8.3cm]{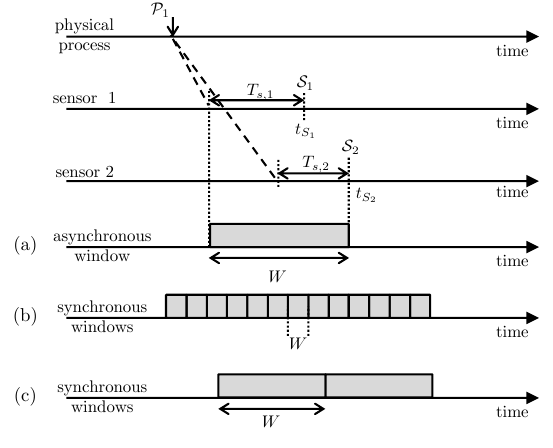}
 \caption{A physical event that is detected by two asynchronous sensors. (a) Use of asynchronous temporal window of integration that starts with the start of the detection of the first sensor. (b) Synchronous windows of integration with $W<t_{S_2}-t_{S_1}$. (c) Synchronous windows of integration with $W \geq t_{S_2}-t_{S_1}$.}
 \label{fig:TwoSensors}
\end{figure}

At first we consider two sensory inputs, with integration times $T_{s1}$ and $T_{s2}$, respectively. The arrival times of the sensory events are:
\begin{eqnarray}
    t_{S_i}=\tau_{s,i}+T_{s,i}
\end{eqnarray}
where $i \in \{1,2\}$. The simplest case is the one of asynchronous sensors, where the detection starts at the instant at which the sensory event becomes detectable, as depicted on Fig.~\ref{fig:TemporalOrdering}(b). This is reminiscent of multisensory perception in humans for, say, visual and audio inputs~\cite{spence2003multisensory}, and, following this line of thought, we want to find the condition for selecting the window size $W$ that guarantees correct simultaneity/synchrony of the received events. 

Fig.~\ref{fig:TwoSensors} illustrates detection of a physical event with two sensors. Fig.~\ref{fig:TwoSensors}(a) shows an asynchronous TWI and, without loss of generality, assume that $\tau_{s,1}<\tau_{s,2}$, such that sensor $1$ gets activated first. In order to perceive the two events ${\cal S}_1$ and ${\cal S}_2$ as simultaneous, the following two conditions have to be met:
\begin{equation}
    W  \geq T_{s,1} \qquad \tau_{s,1}+W \geq \tau_{s,2}+T_{s,2}
\end{equation}
leading to:
\begin{equation}
    W = \max\{T_{s,1},\tau_{s,2}-\tau_{s,1}+T_{s,2} \} 
\end{equation}
To make a parallel with human audio-visual perception, assume that sensor 1 is visual with $T_{s,1}=50$ ms and sensor 2 is acoustic with  $T_{s,2}=10$ ms \cite{vroomen2010perception}. This is consistent with $\tau_{s,1}<\tau_{s,2}$ as light propagates faster than sound. The human window of integration is chosen such that simultaneity is preserved within $10-15$ m, also termed \emph{horizon of simultaneity}~\cite{poppel1990sensory}. This indicates that the TWI for asynchronous digital sensors has to be chosen to satisfy the simultaneity requirement within a certain spatial setup. Regarding the event throughput loss, if $T_{s,2}<T_{s,1}$ then the loss is:
\begin{equation}
    \frac{W}{T_{s,2}} \geq \frac{\tau_{s,2}-\tau_{s,1}+T_{s,2}}{T_{s,2}}=1+\frac{\tau_{s,2}-\tau_{s,1}}{T_{s,2}}
\end{equation}
that is, the propagation times have an impact on the event throughput loss. 

Consider now synchronous TWI, as in Fig.~\ref{fig:TwoSensors}(b) and (c). 
If the TWI is of a size $W < (t_{S_2}-t_{S_1}))$, then simultaneity violation occurs with probability $P({\cal V}_s)=1$ since, as seen from Fig.~\ref{fig:TwoSensors}(b), the events ${\cal S}_1$ and ${\cal S}_2$ always have different timestamps, $s(t_{S_1})<s(t_{S_2})$. If $W \geq t_{S_2}-t_{S_1}$ then simultaneity violation occurs when an edge of the synchronous TWIs falls between ${\cal S}_1$ and ${\cal S}_2$. Assuming that the start of the synchronous windows is chosen uniformly the probability of simultaneity violation is $P({\cal V}_s)=\frac{t_{S_2}-t_{S_1}}{W}$. In summary, this can be written as:
\begin{equation}
    \Pr[{\cal V}_s]=\min\left\{ 1, \frac{t_{S_2}-t_{S_1}}{W} \right\}
\end{equation}

\subsection{Causality Violation with a Sensing and a Digital Input}

The main difference introduced by the digital input is the variable duration of the reception period $T_d=\frac{D_d}{R}$ which depends both on the data size and the data rate. Both $D_d$ and $R$ are, at least partially, subject to choice by the sender $A$, which can have an impact on the causality/simultaneity of the received events. For instance, $A$ can choose $D_d$ to be smaller and thus represent a certain source in a coarser, more distorted way. It thus becomes interesting to consider the case in which at least one of the events at the receiving device $B$ is digital due to the possibility to have an external entity that has an impact on the rate of information processing, unlike content of the sensory inputs, decided by physics/nature.

\begin{figure}[t!]
 \centering
 \includegraphics[width=8.3cm]{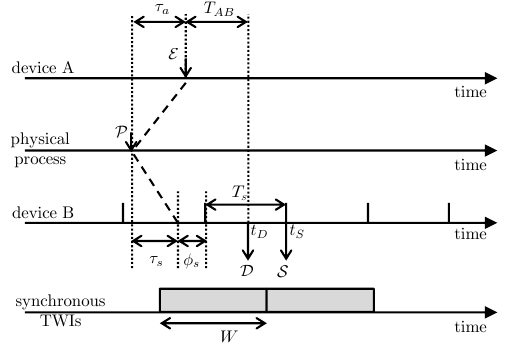}
 \caption{Physical event triggers a digital transmission. Example of causality violation at device $B$ when the event ${\cal D}$ of decoding the data from $A$ precedes the sensing event ${\cal S}$ and the events are timestamped such that $s(t_{D})=s(t_S)-1$.}
 \label{fig:AnalogCausesDigital}
\end{figure}

In this subsection we consider the temporal ordering and causality violation at a device $B$ with a synchronous sensor and a digital receiver; that is, $B$ observes an integrated sensing and digital process. Two different cases are treated. In the first case, depicted on Fig.~\ref{fig:AnalogCausesDigital}, a physical event triggers a digital data transmission at device $A$. In the second case from Fig.~\ref{fig:DigitalCausesAnalog}, device $A$ triggers the physical event and at the same time it transmits digital data to $B$; it can be interpreted as the digital transmission triggers a physical event that is detected by $B$'s sensor. Regarding the synchronous \ac{twi}, it is defined and run at a higher system levels. 

\subsubsection{Case 1: Physical Event Triggers A Digital Transmission}

From Fig.~\ref{fig:AnalogCausesDigital}, the physical event ${\cal P}$ arrives at the sensor $\phi_s \in [0,T_s)$ seconds before the start of a sensor integration window. Recall that, in our model, the event is not detected reliably if the sensor collects the data for time $\phi_s < T_s$. This, in combination with the assumption about predefined sensor integration windows, leads to the detection of the physical event as a sensing event ${\cal S}$ at the end of the subsequent sensor integration window. The event ${\cal P}$ causes a digital transmission event at device $A$ after time $\tau_a$ and the data is received by device $B$ after time $T_{AB}$, defined in (\ref{eq:TotalTimeDigitalEvent}), resulting in a digital event ${\cal D}$. The times at which the sensing and the digital event are registered are:
\begin{equation} \label{eq:defntDtS1}
    t_D=\tau_a+T_{AB} \qquad t_S=\tau_s+\phi_s+T_s
\end{equation}
Considering the timestamping with the TWI, causality violation occurs if at device $B$ the event ${\cal D}$ happens before ${\cal S}$, that is $s(t_D)<s(t_S)$. Following a similar line of thought as in the case with two sensors, causality violation occurs if $t_D<t_S$ and there is at least one TWI edge between $t_D$ and $t_S$. Assuming that the TWIs are not synchronized with the sensing or digital process, then the probability of causality violation can be expressed as follows: 
\begin{equation}\label{eq:PCV-analog-digital}
    \Pr[{\cal V}_c]=\min\left\{ 1, \frac{\max\{0,t_S-t_D\}}{W} \right\}
\end{equation}
Causality violation never occurs if $t_S-t_D<0$. Substituting the expressions from (\ref{eq:defntDtS1}) this becomes:
\begin{equation}\label{eq:phiscondition1PHYDIG}
    \phi_s<T_{AB}+\tau_a-T_s-\tau_s
\end{equation}
Since $\phi_s<T_s$ it follows that causality violation never occurs if the right-hand side of (\ref{eq:phiscondition1PHYDIG}) is larger than $T_s$, or equivalently:
\begin{equation}\label{eq:zeroviolation1PHYDIG}
    T_{AB}>2T_s+\tau_s-\tau_a
\end{equation}
Differently from this, causality violation is certain whenever $t_S-t_D \geq W$ which can be written as:
\begin{equation}\label{eq:phiscondition2PHYDIG}
    \phi_s \geq W+T_{AB}+\tau_a-T_s-\tau_s
\end{equation}
Since $\phi_s$ is nonnegative, the condition for certain causality violation can be written as:
\begin{equation}\label{eq:certainviolation1PHYDIG}
   W<T_s+\tau_s-T_{AB}-\tau_a
\end{equation}
On the other hand, to be able to mitigate the causality violation, from (\ref{eq:phiscondition2PHYDIG}) it follows:
\begin{equation}
    W \geq T_s+\tau_s+\phi_s-T_{AB}-\tau_a
\end{equation}
In order to be satisfied for all values of $\phi_s, T_{AB}, \tau_s,$ and $\tau_a$, the minimal window size is:
\begin{equation}\label{eq:WminPHYDIG}
    W_{\min} \geq 2T_s+\tau_{s,\max}-T_{\min}-\tau_{a,\min}
\end{equation}
where it is set $\phi_s=T_s$ and $T_{AB}=T_{\min}$, assuming that $T_{AB}$ is a random variable with a support set in $T_{\min},T_{\max}$, while $\tau_{s,\max}$ and $\tau_{a,\min}$ are the maximal value of $\tau_s$ and minimal value of $\tau_a$ in a certain physical-computational setup. Note that this does not guarantee that causality violation will not occur, the choice  $W \geq W_{\min}$ only makes the probability of causality violation to be less than one.

\noindent \textbf{Example 1.} Consider a wireless cellular system where the device $B$ is a Base Station that uses integrated communication and sensing. Assume that both the physical process, where ${\cal P}$ happens, and the device $A$ are at a distance of $3$ km from device $B$. Assume that the distance between device $A$ and ${\cal P}$ is $30$ m and the effect of ${\cal P}$ propagates to $A$ at the speed of sound. Then $\tau_a \approx 100$ ms, while $\tau_s = \tau_d = 10$ $\mu$s, as both are based on radio propagation. Let $T_s$ be in the range of ms. If ${\cal P}$ causes the digital transmission of $A$, then for these values we have that $2T_s+\tau_s-\tau_a<0$, such that (\ref{eq:zeroviolation1PHYDIG}) is always satisfied and causality violation does not occur. However, assume that $A$ can accurately predict ${\cal P}$ and or generate a fake data to be transmitted; this can decrease $\tau_a$ significantly and may lead to causality violation.

\subsubsection{Case 2: Digital Transmission Triggers a Physical Event} 

\begin{figure}[t!]
 \centering
 \includegraphics[width=8.3cm]{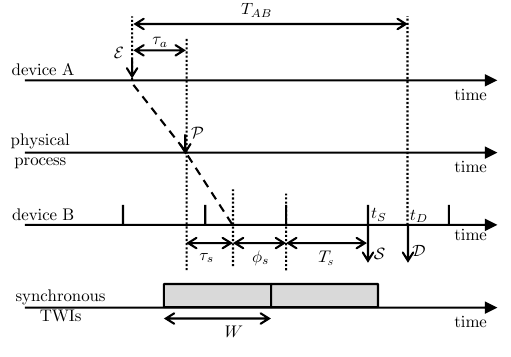}
 \caption{Digital transmission triggers a physical event. Example of causality violation at device $B$ when the sensing event ${\cal S}$ is detected before the digital data arrives at time of ${\cal D}$.}
 \label{fig:DigitalCausesAnalog}
\end{figure}

As Fig.~\ref{fig:DigitalCausesAnalog} depicts, the event ${\cal E}$ at device $A$ causes the physical event ${\cal P}$. At the same time with ${\cal E}$, the device $A$ starts the transmission of a data packet and $B$ receives it after time $T_{AB}$, corresponding to ${\cal D}$. On the other hand, the sensing event ${\cal S}$ is detected before ${\cal D}$, which is a causality violation, since it may lead to conclusion that ${\cal P}$ happened before, and even caused, ${\cal E}$. The times at which the sensing and the digital event are registered are now expressed differently:
\begin{equation} \label{eq:defntDtS2}
    t_D=T_{AB} \qquad t_S=\tau_s+\tau_a+\phi_s+T_s
\end{equation}

With timestamping, causality violation occurs when $s(t_S)<s(t_D)$. Using the same arguments as in the previous section, the probability of causality violation is found to be:
\begin{equation}\label{eq:PCV-digital-analog}
    \Pr[{\cal V}_c]=\min\left\{ 1, \frac{\max\{0,t_D-t_S\}}{W} \right\}
\end{equation}
Causality violation never occurs if $t_D-t_S<0$, leading to:
\begin{equation}\label{eq:phiscondition1DIGPHY}
    \phi_s>T_{AB}-\tau_a-T_s-\tau_s
\end{equation}
Since $\phi_s \geq 0$, causality violation is avoided if:
\begin{equation}\label{eq:nocvDIGPHY}
    T_{AB}=T_{\max}<T_s+\tau_a+\tau_s
\end{equation}
Causality violation occurs with probability $1$ if:
\begin{equation}\label{eq:certcvDIGPHY}
    W < T_{AB}-T_s-\tau_a-\tau_s-\phi_s
\end{equation}
Hence, the minimal size of the TWI required to be able to mitigate the causality violation is obtained for $T_{AB}=T_{\max}$ and $\phi_s=0$:
\begin{equation}\label{eq:WminDIGPHY}
    W_{\min} \geq T_{\max}-T_s-\tau_{a,\min}-\tau_{s,\min}
\end{equation}

\noindent \textbf{Example 2.} In this case $B$ is a LEO satellite, while ${\cal P}$ and $A$ are on the ground and in proximity of each other. In this case $\tau_s$ and $\tau_d$ are in the order of milliseconds, while $\tau_a$ depends on how $A$ causes detectable changes expressed by ${\cal P}$. Let $\tau_a$ be in the order of milliseconds and $T_s$ at the satellite around $100$ ms. If the data payload that $A$ sends to $B$ is large, while the data rate $R$ is low due to the low link budget, then causality violation can easily happen according to (\ref{eq:nocvDIGPHY}). 

\noindent \textbf{Example 3.} Assume that device $A$ has an accurate learning model of the physical process and can predict the event ${\cal P}_1$. In that case, $\tau_a=0$ or it can even be negative (!), increasing the chances of causality violation in the case when in reality ${\cal P}_1$ should happen before and cause the transmission of $A$.  


\section{Cases with $N>2$ Inputs}
\label{sec:Ninputs}

\begin{figure}[t!]
 \centering
 \includegraphics[width=8.3cm]{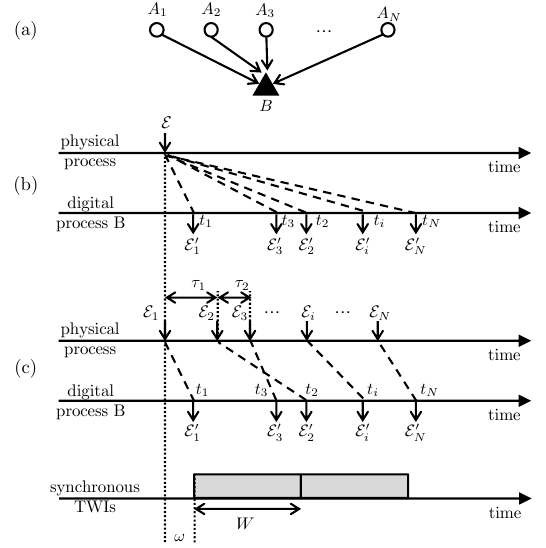}
 \caption{(a) Setup with $N$ different senders $\{A_i\}$ providing inputs to a common receiver $B$. (b) Case for analysis of simultaneity violation, when a single event results in $N$ different digital data receptions. (c) Case for analysis of causality violation. When no TWI is applied, there is causality violation between ${\cal E}^{\prime}_2$ and ${\cal E}^{\prime}_3$. The synchronous TWIs is plotted for both (b) and (c).}
 \label{fig:CaseNmore2}
\end{figure}

We now consider a setup in which the receiver has $N$ different inputs. Two cases are considered: (1) Simultaneity violation upon a single event that initiates $N$ transmissions; (2) Causality violation for $N$ causally linked events, each of them resulting in a reception of digital data.  
The setup is given in Fig.~\ref{fig:CaseNmore2}(a), where $N$ different devices $\{A_i\}$ transmit data to a common destination/server $B$. Note the definition of the TWI offset $\omega$, uniformly distributed in $[0,W)$ with respect to the sole event on Fig.~\ref{fig:CaseNmore2}(b) and with respect to the first event on Fig.~\ref{fig:CaseNmore2}(c).

\subsection{Simultaneity Violation}

Referring to~Fig.~\ref{fig:CaseNmore2}(b), a single physical or digital event ${\cal E}$ causes reception of $N$ events
${\cal E}^{\prime}_1, {\cal E}^{\prime}_2, \ldots, {\cal E}^{\prime}_N$ at times $t_1, t_2, \ldots, t_N$, respectively. The model is easily applicable to both digital and sensing events. Assuming that the event ${\cal E}$ happened at $t=0$, then, if the $i-$th event is a digital one ${\cal E}^{\prime}_i={\cal D}_i$ we have:
\begin{equation}
    t_i=\tau_{a,i}+T_{A_iB}
\end{equation}
Otherwise, if the $j-$th event is a sensing event, then  ${\cal E}^{\prime}_j={\cal S}_j$ and:
\begin{equation}
    t_j=\tau_{s,j}+T_{s_j}
\end{equation}
Simultaneity violation does not occur if the following is satisfied:
\begin{equation}
    s(t_1)=s(t_2)=\cdots=s(t_N)
\end{equation}
Let us assume that synchronous TWIs of size $W$ that are not synchronized with the arriving events. Given the set of detected event times ${\cal T}={t_1, t_2, \ldots t_N}$ and assuming TWI offset $\omega$ is uniformly distributed in $[0,W)$, the probability of simultaneity violation is:
\begin{equation}\label{eq:PVs--T}
    \Pr[{\cal V}_s|{\cal T}]=\min\left\{ 1, \frac{t_{\max}-t_{\min}}{W} \right\}
\end{equation}
where $t_{\min},t_{\max}$ are the minimal/maximal values in ${\cal T}$, respectively. Given the probability distribution $P({\cal T})$, one can calculate the total probability of causality violation. It is reasonable to assume that the times $T_{A_iB}$ are conditionally independent, given the occurrence of the event that caused them. Based on this, all $t_i$ are independently distributed, such that $P({\cal T})$ is a product distribution $p_1(t_1)\cdot p_2(t_2) \cdots p_N(t_N)$. However, note that, if the $i-$th input is from an asynchronous sensor, then $\phi_{s_i}=0$ and there is no randomness in $t_i$. A more general analysis would consider joint distribution $p_1(t_1,t_2, \cdots t_N)$, in which $T_{A_iB}$ are still dependent, even when conditioned on the event ${\cal E}$, for example, due to the nature of the data generated by the event. This general case is, though, out of the scope for this manuscript.      

\subsection{Causality Violation with Window Size $W=0$}

Fig.~\ref{fig:CaseNmore2}(c) shows the setup for analysis of causality violation with $N$ inputs. To start with, we assume that all $N$ inputs are digital such that ${\cal E}^{\prime}_i={\cal D}_i$ for all $i$; we will revise this assumption later on. There is a temporal/causal ordering of the physical-world events $\{{\cal E}_i\}, i \in \{1, 2, \ldots N\}$, that result in digital transmissions. The event ${\cal E}_i$ causes the event ${\cal E}_{i+1}$ after time $\tau_i$, where $i=1, \ldots N-1$. We consider the random time $T_i$ it takes for $B$ to receive ${\cal E}_i$ from device $A_i$ as an event ${\cal D}_i$. The event ${\cal E}_1$ occurs at time $t=0$ while the event ${\cal D}_i$ is received at time (within the reference frame of the physical process):
\begin{equation}\label{eq:definitionti}
    t_i=T_i+\sum_{k=1}^{i-1}\tau_k
\end{equation}
for $i>1$ and $t_1=T_1$. 

In this section we treat the case without a TWI, which corresponds to $W=0$. Since ${\cal E}_i \rightarrow {\cal E}_j$ whenever $i<j$, there is no causality violation if the following is satisfied:
\begin{equation}\label{eq:NoCVNconditionSW0}
    t_1 \leq t_2 \leq \cdots \leq t_N
\end{equation}
Let use define the indicator function:
\begin{equation}\label{eq:IndicatorDEFN}
    I(t_i\leq t_j)=I_{i,j}=
        \left\{\begin{array}{cc}
                1 & \text{if }t_i \leq t_j\\
                0 & \text{otherwise}
        \end{array}\right.
\end{equation}
We define the event 
$I_{i,j}=1$ as the correct causal ordering ${\cal D}_i \rightarrow {\cal D}_j$ whenever $i<j$:
For a set of $N$ events, there is no causality violation if $I_{i,j}=1$ holds for any pair $i<j$. Considering the ordering of the events in the physical process, when there is no causality violation the ordering of the digital events is 
${\cal D}_1 \rightarrow {\cal D}_2 \rightarrow \ldots {\cal D}_N$. Since causality is a transitive relation, this implies that ${\cal D}_i \rightarrow {\cal D}_j$ when $i>j$. Hence, the probability of not having a causality violation is simplified to:
\begin{equation}\label{eq:NOCVsuccessiveEventsW0}
\Pr\left[\prod_{i<j} I_{i,j} = 1 \right]=\Pr\left[\prod_{k=1}^{N-1} I_{k,k+1} =1\right]
\end{equation}
Considering the definition (\ref{eq:definitionti}), it can be seen that:
\begin{equation}\label{eq:Ckk+1Event}
    I_{k,k+1} =1 \Longleftrightarrow t_k \leq \tau_k+t_{k+1}
\end{equation}
We proceed towards deriving an upper bound on the probability (\ref{eq:NOCVsuccessiveEventsW0}) for correct causal ordering of $N$ received events. The main idea is to decouple the pairwise probabilities since, in general, $\Pr\left[\prod_{k=1}^{N-1} I_{k,k+1} =1 \right]\neq \prod_{k=1}^{N-1} \Pr[I_{k,k+1} =1]$ since different events $I_{k,k+1}=1$ are dependent as they share common random variables. To find the upper bound, at first we need the following lemma. 
\begin{lemma}
    Let $t_1, t_2,$ and $t_3$ be independent random variables. Then:
    \[
    \Pr\left[t_2 \leq t_3|t_1 \leq t_2\right] \leq \Pr\left[t_2 \leq t_3\right]
    \]
\end{lemma}
The intuition behind this is that it is more probable for 
$t_2$ to be less or equal to $t_3$ when $t_2$ is unrestricted, rather than when it is known that $t_2$
is larger than another value. The proof is given in Appendix~\ref{sec:ProofLemma1}.

We can now show that the probability of correct causal ordering is upper bounded by the product of the probabilities for correct causal ordering of neighboring events; the proof is given in Appendix~\ref{sec:ProofTheorem1}.
\begin{theorem}\label{th:Theorem1}
    The probability of correct causal ordering of $N$ digital events, without using temporal integration window, is upper-bounded as follows:
    \[
    \Pr\left[\prod_{k=1}^{N-1} I_{k,k+1}=1\right] \leq \prod_{k=1}^{N-1} \Pr\left[I_{k,k+1}=1\right] 
    \]
\end{theorem}
    
The result is valid if a subset of the inputs are sensing instead of digital links. If $i$ is a sensing input ${\cal E}^{\prime}_i={\cal S}_i$, then $T_i=\phi_{s,i}+T_{s,i}$ and the distribution of $T_i$ can be derived from the distribution of $\phi_{s,i}$. However, for the proof of the theorem to be valid, we need to assume that each input comes to a different sensor. To see why this is the case, assume that $i,j$ with $j>i$ are two sensing inputs. Then, given the set of values $\{\tau_k\}$, the random variables $\phi_{s,i}$ and $\phi_{s,j}$ are not independent, as they are related to the start of the synchronous sensor integration times of the same sensor. 

If the occurrence times of the original events $\{{\cal E}_k\}$ are chosen randomly, resulting in random $\{\tau_k\}$, then the problem can be posed as a problem of concomitants in order statistics~\cite{david2004order}. As a general statement, what we perceive or register with instruments is not the ordering of the original events, but their concomitants.   

We reiterate the justification of considering causality violation for digital inputs. The devices $\{A_i\}$ may have perfectly synchronized clocks and add timestamps to the transmitted data, such that no causality violation can occur. This is only seemingly correct, since these timestamps are added at a certain point within the digital devices and are concomitants for the true order in which the events caused each other in the physical world.

\subsection{Causality Violation with Window Size $W>0$}

By applying TWIs, the condition for no causality violation turns into:
\begin{equation}\label{eq:NoCVNconditionSW>0}
    s(t_1) \leq s(t_2) \leq \cdots \leq s(t_N)
\end{equation}
where $t_i$ is defined in (\ref{eq:NoCVNconditionSW0}). 
Note that once the random TWI offset $w$ is chosen, it is identical for all the detected events. We thus cannot use the proof of Theorem 1, which relies on the fact that $\Pr\left[ I_{k,k+1}=1   \left\vert \prod_{l=1}^{k-1} I_{l,l+1} = 1\right.\right]=\Pr[t_k \leq t_{k+1} | t_{k-1} \leq t_{k}]$. This is not necessarily valid when we consider $s(t_k)$ instead of $t_k$. We therefore derive a different upper bound for the case $W>0$. With a slight abuse of notation, in this section we use $I_{i,j}$ as $I(s(t_i),s(t_j))$ rather than $I(t_i,t_j)$.    
\begin{theorem}
    The probability of correct causal ordering of $N$ digital events with temporal integration window of size $W>0$, is upper-bounded as follows:
    \[
    \Pr\left[\prod_{k=1}^{N-1} I_{k,k+1}=1\right] \leq \prod_{k=1}^{N-1} \left(\Pr\left[I_{k,k+1}=1\right]\right)^{\frac{1}{N-1}} 
    \]
\end{theorem}
The proof relies on the use of H\"older's inequality~\cite{rudin1976principles} and is given in Appendix~\ref{sec:ProofTheorem2}. Let $P_{\max}$ denote the maximal probability $\Pr\left[I_{k,k+1}=1\right]$ for all $k$; then the upper bound becomes rather trivial:
\begin{equation}\label{eq:UpperNoCVN}
    \Pr\left[\prod_{k=1}^{N-1} I_{k,k+1}=1\right] \leq P_{\max}^{\frac{N-1}{N-1}}=P_{\max}
\end{equation}
which is intuitively correct. 

\section{Numerical Illustration}
\label{sec:NumIllustration}

\begin{figure}[t!]
 \centering
 \includegraphics[width=8.3cm]{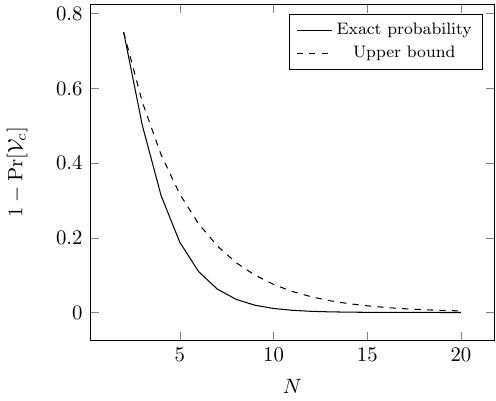}
 \caption{Illustration of the probability of not having causality violation $1-\Pr[{\cal V}_c]$ and its upper bound when the digital transmitters can use two different rates $\frac{R_0}{2}, R_0$ with equal probability.}
 \label{fig:drates}
\end{figure}

We start by a simple verification of Theorem~\ref{th:Theorem1}, in which $W=0$. Assume that there is a sequence of $N$ causally related events ${\cal E}_1,{\cal E}_2, \ldots {\cal E}_N$, as in Fig.~\ref{fig:CaseNmore2}(c), where the action times are identical $\tau_1=\tau_2=\cdots=\tau_{N-1}=\tau$. Next, each device $A_i$ can use only two discrete rates, low rate $\frac{R_0}{2}$ and high rate $R_0$ and, for simplicity, the transmission time corresponding to the low rate is $T=2T_0$, while the one for the high rates is $T=T_0$. We say that a device $A_i$ is in a state $S_i=L/H$ if it sends at a low/high rate respectively. Assume that $T_0>\tau$; then the causality for events  ${\cal E}_k,{\cal E}_{k+1}$ is violated if the states of $A_k$ and $A_{k+1}$ are $S_kS_{k+1}=LH$. For all other possibilities $S_kS_{k+1} \in \{LL,HL,HH\}$ it is easy to see that there is no causality violation. Assuming that the probability of $L/H$ state is $\Pr[L]=\Pr[H]=\frac{1}{2}$ then the probability of no pairwise causality violation, for all $k=1, \ldots N-1$ is:
\[
\Pr[I_{k,k+1}=1]=\Pr[S_kS_{k+1} \in \{LL,HL,HH\}]=\frac{3}{4}
\]
Looking at the states of the entire sequence of transmitters, it can be noted that there a causality violation occurs if there is an $i$ for which $S_i=L$, then there exists $j>i$ for which $S_j=H$. To see why this is the case, note that there must be at least one $i\leq i <j$ for which the state sequence is $S_{l}S_{l+1}=LH$, which leads to causality violation. Hence no causality violation occurs if the entire sequence of states $S_1S_2 \cdots S_N$ takes values from the following sequences:
\[
LLL\ldots LL, HLL\ldots LL, HHL\ldots LL, \cdots
\]
\[
\cdots, HHH\ldots HL, HHH\ldots HH
\]
Since $\Pr[L]=\Pr[H]=\frac{1}{2}$, then the probability of not having causality violation is:
\[
1-\Pr[{\cal V}_c]=\Pr\left[\prod_{k=1}^{N-1} I_{k,k+1} =1\right]=(N+1)\frac{1}{2^N}
\]
From Theorem~\ref{th:Theorem1}, the upper bound on this probabilities based on pairwise probability $\frac{3}{4}$ is:
\[
\prod_{k=1}^{N-1} \Pr\left[I_{k,k+1}=1\right]=\left(\frac{3}{4}\right)^{N-1}
\]
Fig.~\ref{fig:drates} shows the relation between the upper bound and the exact value of the probability of correct causal ordering, that is, not having causality violation, denoted as $1-\Pr[{\cal V}_c]$. What is not visible on the figure is that the the upper bound becomes looser as $N$ grows; that is, the ratio between the upper bound and the exact probability, given as $\left(\frac{3}{2}\right)^{N-1}\frac{2}{N+1}$ is monotonously increasing. This is because, as $N$ grows there are more combinations that can lead to causality violation and they are not captured by the pairwise probabilities. 

\begin{figure}[t!]
 \centering
 \includegraphics[width=8.3cm]{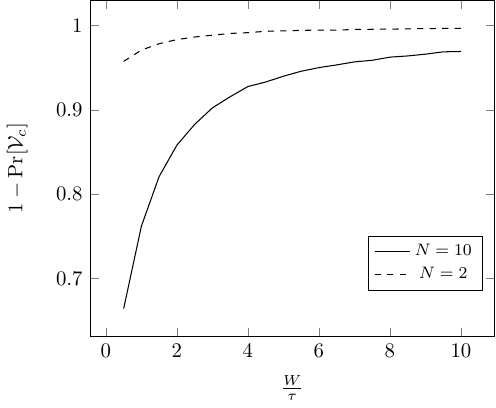}
 \caption{Probability of correct causal ordering for $N=10$ events, where the action time between consecutive events is $\tau$, the TWI size is $W$ and the transmission time for each event is exponentially distributed with $\bar{T}=0.5\tau$. The pairwise probability of correct causal ordering, given by $N=2$, acts as an upper bound for the case $N=10$.}
 \label{fig:PrNoCVN2N10}
\end{figure}

Next we examine the impact of the TWI size on causality violation, with the results shown on Fig.~\ref{fig:PrNoCVN2N10}. We consider $N$ casually related events, where the action time between two events is constant and equal to $\tau$. The transmission times $T_i$ are i.i.d. random variables with exponential distribution that has a mean value $\bar{T}=0.5\tau$. The $x-$coordinate of Fig.~\ref{fig:PrNoCVN2N10} shows $\frac{W}{\tau}$, which corresponds to the event throughput loss. Namely, if the transmission times $\{T_i\}$ are constant, then the event throughput would be $\frac{1}{\tau}$, with a correct causality order; when the window of size $W$ is used, then the event throughput is $\frac{1}{W}$. The curve $N=10$ shows the probability of not having causality violation when all $N=10$ transmissions are considered jointly, that is, $\Pr\left[\prod_{k=1}^{9} I_{k,k+1}=1\right]$. The curve $N=2$ is an upper bound, rather loose, on the pairwise probability of no causality violation, according to (\ref{eq:UpperNoCVN}), since, due to symmetry, here we have that $\Pr\left[I_{k,k+1}=1\right]$ is constant for all $k$. Further analysis of the probability of pairwise probability of correct causal ordering, in a more general setting, is given in Appendix~\ref{sec:UpperBoundPairwise}.

\section{System-Level Perspectives}
\label{sec:SystemLevel}

In this section we look into the implications that the timestamping has on the requirements and design of wireless communication systems, integrated with sensing. 

\subsection{Setting the Communication Latency Requirements}

Latency requirements have been the focus of mobile communication systems since 5G. Some of the earliest requirements set in 5G pointed out to a latency of $1$ ms, while very often 6G is associated with even lower latencies (in $\mu s$!). As discussed in details in~\cite{popovski2022perspective}, putting such maximalist requirements on the radio links is neither always feasible nor can be efficiently supported, such that we need to look into a set of more general timing requirements. In this context, TWI can be used as a universal interface between a real-time application and a set of digital/sensory inputs, such that the latency requirements of a digital link are decided in synergy with the other inputs and, eventually, the TWI and the timestamping requirements. 

Let us assume that, in a given setting, we have $\tau_s-\tau_a=20$ ms, while $T_s=10$ ms. If we want to avoid causality violation when the digital causes a sensing event, from (\ref{eq:nocvDIGPHY}) it follows that the highest allowable value for $T_{AB}$ is $T_s+\tau_a+\tau_s=30$ ms. However, from the definition (\ref{eq:TotalTimeDigitalEvent}) it follows that these $30$ ms include processing at the sender $T_A$, propagation time (which may be negligible) and the actual transmission time $T_B$. Putting some timing constraints on the sender to, say, process the data within $T_A<15$ ms, we have a latency requirement for the radio link to be $15$ ms.    

As a next example, assume that an application has selected the size of the timestamping window $W$ and, based on that, we need to determine the required latency for communication. One way in which the application can choose $W$ is based on a control loop or sense-compute-actuate cycle, where $W$ is determined based on which inputs are required to be able to do the necessary inference and actuation. Note that different sensing and digital input may belong to different systems that are not necessarily synchronized; we can thus put the latency constraint in a way that satisfies a certain probability of causality or simultaneity violation. We want to see how to set the latency requirement for the transmission time $T$ by the digital transmitter $A$ based on the edge of the current TWI. We consider two cases regarding the synchronization of $A$ with respect to TWI:
\begin{itemize}
    \item $A$ knows the start of the TWI and prepares its transmission accordingly. Then a violation in causality/simultaneity occurs if $A$'s transmission does not get through by the end of the window. If the probability of causality violation should not be larger than $\epsilon$, then we can design the resource allocation and the transmission to satisfy $\Pr[T>W] \leq \epsilon$.
    \item $A$ does not know the start of the TWI, that is, $\omega$ is uniformly distributed in $[0,W)$. Let $p(t)$ denote the probability density function of $T$. For given $T=t$, the probability of not sending before the TWI edge equals the probability that the transmission starts later than $t$ seconds before the end of the current window, which is $\frac{t}{W}$. Then the probability of violation of the timestamp is found as:
    \begin{equation}
        \Pr[T>W]=\int_0^{\infty} \frac{t}{W} p(t) \mathrm{d}t=\frac{E[T]}{W}
    \end{equation}
    where $E[T]$ is the expected value of $T$. For instance, let $T_s=10$ ms and let the event throughput loss be $3$, such that $W=3\cdot 10=30$ ms. If $W=30$ ms then even with $T=3$ ms the probability of violation of the timestamp is still $0.1$, that is, we have a little control over it. 
\end{itemize}
From the above examples it follows that, when TWI is used, the better the knowledge that the transmitter has about the start of the window, the better the guarantees for  latency/timestamp violation. In this sense, the uniform distribution of $w$ is the worst case; one can think of algorithms ran by the devices for probabilistic tracking of the start of the TWIs, which would lead to better results. 

Regarding simultaneity violation, an illustrative case is given by a random access protocol. Assume that a physical event happens, upon which $N$ devices detect quantities related to the event and want to transmit to the BS. The $i-$th device detects the event after time $\tau_{a,i}$ and generates a data packet that is transmitted using a random access protocol. In addition, the BS has a sensor that can detect the event after it propagates for $\tau_s$ seconds and the sensing time is $T_s$. A protocol that is particularly suitable for batch arrival is frameless ALOHA~\cite{stefanovic2012frameless}. This is a slotted protocol, with slots of duration $T_1$ and after $M$ slots the fraction of resolved users is $P_R$, such that the throughput is $\frac{NP_R}{MT_1}$. In the original proposal~\cite{stefanovic2012frameless}, the protocol execution is stopped when the throughput is maximal. In the considered variant with sensing, the protocol can be stopped after the window of simultaneity has expired; given the parameters $\{\tau_{a,i}\}, \tau_s, T_s$ and $T_1$, the size of the simultaneity window can be chosen in a way that can capture the desired fraction of the transmissions. Interestingly, using the sensing information some of the transmissions can be predicted, such that this can be used as a side information in the process of collision resolution and decoding. 

\subsection{ISAC and Timestamping Base Stations}
\label{sec:ISACtimestamping}

\begin{figure}[t!]
 \centering
 \includegraphics[width=8.3cm]{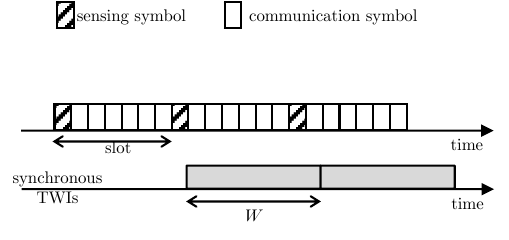}
 \caption{Synchronous slot/symbol structure in 5G NR with time-multiplexed sensing and communication slots. The frequency division is not represented as we are focused on the timing aspects.}
 \label{fig:ISACframe}
\end{figure}

We now look into ISAC, in which the BSs use radar sensing. This is seen as one of the main developments in 6G~\cite{liu2022integrated,wild2021joint}. There are multiple design issues regarding the required spectral resources for radar sensing based on the targeted scenarios, interference between communication and sensing, or joint waveforms for resource allocation and sensing. Yet, from the perspective of the model used in  work, the most important change that ISAC brings is the fact that communication and sensing are time-multiplexed within a synchronous frame used by the mobile system, with a certain numerology~\cite{wild2021joint}. Fig.~\ref{fig:ISACframe} illustrates this time multiplexing, which brings the following three changes in the model: 
\begin{itemize}
    \item Times $\{t_{\cal{S}},t_{\cal{D}}\}$ become discrete, according to the structure of frame/subframe/slot/symbol. For example, the minimal discretized time unit can correspond to the shortest slot in 5G NR numerology, which is $62.5$ $\mu$s. Considering the frequency dimension, a group of digital transmissions that are carrying transparent sensing data relevant for the same phenomenon can be scheduled simultaneously; that is, they are inherently simultaneous and not only due to the TWI.
    \item Recall that in our previous analysis we have assumed that there is independence between among the sensor integration windows and the digital transmissions. However, in a frame-based system they are subject to scheduling and resource allocation policy, such that they are not independent. In principle, the analysis with the transmission times $T_i$ can be carried over to this case, taking into account the discrete nature of the values they can get. However, this analysis must take into account the dependency created by the time-multiplexed structure. 
    \item Finally, the sensor model we have used is  limited by the assumption that the integration time of the sensor $T_s$ is equal to the period of the sensor. For the example in~\cite{wild2021joint},  
    if there are $64$ sensing beams then the sweeping period is $64$ ms; however, only a fraction of $\frac{1}{64}$ is given to a specific sensing beam.
    This does not correspond to the radar sensing setup with ISAC frame, as depicted in Fig.~\ref{fig:ISACframe}. Note that this can not always be remedied by simply defining different sensor integration time $T_s$ and a period of the sensor integration window $T_{s,w}>T_s$, since the BS may allocate aperiodic resources for sensing, depending on the application needs. 
\end{itemize}

As indicated in the introduction, there are two principal uses of the TWIs, real-time applications and temporal forensics. Having a synchronous frame structure, one may legitimately ask: do we need the TWI and analysis with windowing? The answer regarding the real-time application is yes, since different applications will have different TWIs and the scheduling/resource allocation should be capable to gather the required inputs within the respective windows. In the simplest case, a TWI can be a multiple of the frame/slot size, but this does not need to be the case; the only requirement is that the TWI is a multiple of the smallest discrete unit adopted in the system. In terms of temporal forensics, one may argue that the devices that are synchronized to the BS can timestamp their packets that carry transparent sensing data. However, while this can mitigate causality violation, the true timing of the sensory events remains unknown.  

In order to illustrate the usefulness of TWI, consider a complex setup in which the BS has radar sensing, but also other sensors (e.g. acoustic or visual) and, in addition, gets transparent sensing data through devices connected to it, referred to as passive \ac{tmt} in~\cite{xie2022perceptive,xie2023collaborative}. The works dealing with TMTs mention only ``strict latency requirements'', while the use of TWI established meaningful timing requirements from all sensors and digital transmissions. This leads to interesting problems of scheduling and resource allocation. For instance, the amount of radar sensing resources required within the frames/slots depends on the resolution/precision that needs to be attained~\cite{wild2021joint}. The sensing resources are allocated as a comb, not taking time-adjacent resources in order not to block the channel for low-latency transmissions. However, this is rather restrictive: we may look at the TWI structure and allocate the resources in a way that all relevant data for the application/phenomenon can be correctly timestamped in terms of simultaneity and causality.  

In summary, this part highlights one of the most important function that that future BS will have and that is timestamping. Indeed, one can see the BSs and mobile infrastructure as the entry points that put timestamps on the physical reality as it enters the digital world, thus representing the filter that allows trustworthy and temporally correct mapping of the physical into the digital world.


\section{Discussion and Future Work}
\label{sec:discussion}

As the BS and the mobile infrastructure evolve towards entities with multisensory perception, we expect that there will be increased use of brain-like mechanisms a the mobile edge in order to determine the temporal order of the events. In this paper we have discussed one such mechanism, which is the use of TWIs. In~\cite{vroomen2010perception} this is only the first of the four mechanisms used in determining intersensory synchrony and consistent sequence of events. We list the other three mechanisms, with a comment that relates them to the mobile infrastructure:
\begin{itemize}
    \item \emph{``The brain might be “intelligent” and bring deeply rooted knowledge about the external world into play that allows it to compensate for various external factors''}. This corresponds to the use of causal inference at the edge/cloud, based on the trained ML models, which is used in real time to provide the most plausible temporal ordering beyond what is given by the timestamps. 
    \item \emph{``The brain might be flexible and shift its criterion about synchrony in an adaptive fashion''}. This could correspond to different states of criticality within a set of applications; for instance, if no actuation is required within a short time window, then the TWI could be extended. 
    \item \emph{``In order to reduce gaps, the brain might actively shift the time at which one information stream is perceived toward the other.''} This case could correspond to a compensation that the edge can apply using generative AI in order to replace a packet/stream that has not arrived due to transmission failure. 
\end{itemize}
In the case of mobile infrastructure, the complexity increases as the inputs to multiple BSs/edge servers are orchestrated to represent unified timestamping of the physical world. 

The sensing model that we have used in the paper is quite simplified and limited. We have already seen the limitations in the case of modeling a radar sensing with different beams. Another limitation comes from the idealized assumption about the perfect detection. Once we bring imperfect detection in the model, then a number of interesting questions arise. For instance, a single observation within a sensory integration window may give an unreliable detection of the event; however, the detection gets refined in the subsequent windows, which brings the question: when did the sensory event happen and what timestamp should be given? Throughout the paper we have worked with the premise that the rate of information processing determines the sequence of the events, but imperfect sensory detection implies that we should also include the fidelity of the events in the temporal ordering. To exemplify further this argument, in transparent sensing the sender may decide to send a small data portion of size $D$ to describe the sensory event, e.g. image. Thus the event/image would be detected, but with a high distortion; sending a higher quality image would result in a different temporal instant of detection. 

The consideration of timestamping and ordering of events in the mobile infrastructure links to a rich body of literature on logical clocks in distributed systems~\cite{LamportClocks,baquero2016logical}. 
It would be interesting to see how the ISAC-capable BS and edge intelligence can contribute to the creation of consistent causal histories of sensing events. For instance, the BSs can deliver data to the cloud using \emph{superdense time} after discovering causal relationships among the events with the same timestamp~\cite{lohstroh2023logical}. The fact that the sensing events are unlabelled and can be detected by multiple devices/BSs may lead to conclusions in which two sensing events are regarded as causally related, while the ground truth is that is one and the same event; that is, paradoxically, the sensing event has caused itself. As mentioned throughout the paper, timestamping and inference can be used as tools against malicious actors that intend to create fake events or manipulate the chronology/causal ordering of the events.  

Finally, it is interesting to speculate whether the relation between information processing, temporal ordering, and causality has deeper implications in physics. Consider the following thought experiment. A digital device $A$ is positioned at a center point of a circle in space and another digital device $B$ is on a rocket that moves along the circle with high, relativistic speeds. Due to time dilation, the time at $B$ moves at a half pace compared to $A$, that is, while $2$ seconds have passed for $A$, only one second has passed for $B$. Device $A$ perceives the world around it at a data rate $R$ [bps] and has a digital link to $B$. If $B$ encodes the data at a double rate of $2R$ [bps], its data would be received by $A$ at a rate $R$ [bps] and $A$ may conclude that the pace of change in $B$'s world is the same as in the surrounding world.     

\section{Conclusions}

Motivated by the increasing importance of sensing and Integrated Sensing and Communications (ISAC) in the emerging wireless systems, we have considered the problem of  temporal ordering and timestamping of events in a wireless infrastructure. This problem is believed to grow in importance as the Base Stations (BSs) and the mobile infrastructure are entry points for the sensing the physical into the digital world and, vice versa, for actuating commands from the digital into the physical realm. We have defined the problems of causality and simultaneity and have shown how can they be modeled and analyzed. Borrowing from the literature on multisensory perception, we have introduced Temporal Window of Integration (TWI) in the mobile infrastructure as an elementary tool to mitigate the problems of simultaneity/causality violation. The paper has analyzed elementary cases of simultaneity/causality violation and highlighted the main tradeoffs. Furthermore, the paper has provided a system-level perspective that emerges from the timestamping and the use of TWIs and related it to the ISAC-enabled BS. One of the most important takeaways from the paper is that the future Base Stations and Access Points will have a \emph{timestamping functionality} that determines the chronology of the events, simultaneity, and causality and in that way ensure trustworthiness in the physical-world events that are mapped to the digital infrastructure or digital twins. Several potential avenues for future work have been outlined.

\appendices

\section{Proof of Lemma 1}
\label{sec:ProofLemma1}
\begin{proof}
We start by using the law of total probability:
\begin{eqnarray} \label{eq:totalprobability}
    \Pr[t_2 \leq t_3] &=&
    \Pr[t_1 \leq t_2] \Pr[t_2 \leq t_3|t_1 \leq t_2] \notag \\
    &+&\Pr[t_1 > t_2] \Pr[t_2 \leq t_3| t_1 > t_2]
\end{eqnarray}
We define:
\begin{eqnarray}
f(t_1,t_2,t_3)=\Pr[t_2 \leq t_3| t_2 < t_1 ] \notag \\
g(t_1,t_2,t_3)=\Pr[t_2 \leq t_3|t_2 \geq t_1] \notag    
\end{eqnarray}
Note that we have switched the inequalities for the event that is conditioned upon, but the expressions stay the same as in (\ref{eq:totalprobability}). Next, we want to show that:
\begin{align}\label{eq:fminusggt0}
    f(t_1,t_2,t_3) \geq g(t_1,t_2,t_3)
\end{align}
Since the random variables are independent, the probability density function (p.d.f.) is given by:
\[
p(t_1=x_1,t_2=x_2,t_3=x_3)=p_1(x_1)p_2(x_2)p_3(x_3)
\]
where $p_i(x_i)$ is the p.d.f. of $t_i$. 
Then (\ref{eq:fminusggt0}) can be written equivalently as:

{\small
\begin{align}\label{eq:intfminusggt0}
\int_{0}^{\infty}\int_{0}^{\infty} \left[
f(x_1,t_2,x_3) - g(x_1,t_2,x_3)\right] p_1(x_1)p_3(x_3)\mathrm{d}x_1\mathrm{d}x_3 \geq 0
\end{align}
}

where we note that there is no marginalization over $t_2$. 
With $t_2$ remaining as a random variable we get:
\begin{eqnarray} \label{fminusgab}
    f(x_1,t_2,x_3) - g(x_1,t_2,x_3)= \notag \\
    \Pr[t_2 \leq x_2|t_2 < x_1]-\Pr[t_2 \leq x_2 |t_2 \geq x_1]
\end{eqnarray}
If $x_1 \leq x_2$ then $\Pr[t_2 \leq x_2|t_2 < x_1]=1$ and:
\begin{equation}
    f(x_1,t_2,x_3) - g(x_1,t_2,x_3)=1-\Pr[t_2 \leq x_2 |t_2\geq x_1] \geq 0
\end{equation}
If $x_1>x_2$ then $\Pr[t_2 \leq x_2 |t_2 \geq x_1]=0$ and:
\begin{equation}
    f(x_1,t_2,x_3) - g(x_1,t_2,x_3)= \Pr[t_2 \leq x_2|t_2 < x_1] \geq 0
\end{equation}
Since this is valid for all $x_1,x_2$, it follows that (\ref{eq:intfminusggt0}) is satisfied. Now (\ref{eq:totalprobability}) can be lower bounded as:
{\small
\begin{eqnarray}
    \Pr[t_2 \leq t_3]= \notag \\
    \Pr[t_1 \leq t_2]g(t_1,t_2,t_3)+\Pr[t_1 > t_2]f(t_1,t_2,t_3) \notag \\
    \stackrel{(a)}{\geq} \Pr[t_1 \leq t_2]g(t_1,t_2,t_3)+\Pr[t_1> t_2]g(t_1,t_2,t_3) \notag \\
    \stackrel{(b)}{=}g(t_1,t_2,t_3)=\Pr[t_2\leq t_3|t_2\geq t_1] \notag
\end{eqnarray}}
\noindent where (a) follows (\ref{eq:fminusggt0}), (b) is due to $\Pr[t_1 \leq t_2]+\Pr[t_1 > t_2]=1$. This proves the Lemma.
\end{proof}

\section{Proof of Theorem 1}
\label{sec:ProofTheorem1}

\begin{proof}
    The joint probability can be written as follows:
\begin{eqnarray}\label{eq:Theorem1-1}
    \Pr\left[\prod_{k=1}^{N-1} I_{k,k+1}=1 \right]= \notag \\ 
    \Pr \left[I_{1,2}=1\right] \prod_{k=2}^{N-1} \Pr \left[ I_{k,k+1}=1   \left\vert \prod_{l=1}^{k-1} I_{l,l+1}=1 \right. \right]
\end{eqnarray}
For a given set ${\tau_i}$ and the definition (\ref{eq:definitionti}) it follows that the value of $I_{l,l+1}$ depends only on the random variables $t_l$ and $t_{l+1}$. Since all random variables $t_1, t_2, \ldots t_N$ are independent, it follows:
\begin{eqnarray}
    &\Pr& \left[ I_{k,k+1}=1   \left\vert \prod_{l=1}^{k-1} I_{l,l+1} = 1\right.\right] = \notag \\
    &\Pr& \left[ I_{k,k+1}=1   \left\vert I_{k-1,k}=1\right.\right] \notag \\
    = &\Pr& [t_k \leq t_{k+1} | t_{k-1} \leq t_{k}] \notag \\ 
    \stackrel{(a)}{\leq} &\Pr& [t_k \leq t_{k+1} ]=\Pr[I_{k,k+1}=1]
\end{eqnarray}
where (a) follows from Lemma 1. Substituting this result in (\ref{eq:Theorem1-1}) proves the Theorem. 
\end{proof}

\section{Proof of Theorem 2}
\label{sec:ProofTheorem2}

\begin{proof}
    We can write the probability as follows:
    \begin{equation}
        \Pr\left[\prod_{k=1}^{N-1} I_{k,k+1}=1\right] = 
        \int \prod_{k=1}^{N-1} I_{k,k+1} \mathrm{d}\mu
    \end{equation}
where $\mathrm{d}\mu$ is the measure defined as 
\begin{equation}
\mathrm{d}\mu=p(t_1)\mathrm{d}t_1p(t_2)\mathrm{d}t_2 \ldots p(t_N)\mathrm{d}t_N p(\omega)\mathrm{d}\omega 
\end{equation}
as the $\{t_i\} \cup \omega$ are independent random variables. Then:
\begin{eqnarray}
    \int \prod_{k=1}^{N-1} I_{k,k+1} \mathrm{d}\mu \stackrel{(a)}{=} \left\| \int \prod_{k=1}^{N-1} I_{k,k+1} \mathrm{d}\mu  \right\|_1 \stackrel{(b)}{\leq} \prod_{k=1}^{N-1}\left\|I_{k,k+1}\right\|_{N-1} \notag
\end{eqnarray}
where (a) follows from $I_{k,k+1} \geq 0$ and (b) follows from H\"older's inequality. We can now write:
\begin{eqnarray}
   \left\|I_{k,k+1}\right\|_{N-1}=\left( \int I_{k,k+1}^{N-1} \mathrm{d}\mu \right)^{\frac{1}{N-1}}\stackrel{(c)}{=} \left( \int I_{k,k+1} \mathrm{d}\mu \right)^{\frac{1}{N-1}} \notag \\
   \stackrel{(d)}{=} \left( \int I_{k,k+1} p(t_k)p(t_{k+1})p(\omega)\mathrm{d}t_k \mathrm{d}t_{k+1} \mathrm{d}\omega \right)^{\frac{1}{N-1}}
   \notag \\
   \stackrel{(e)}{=} \left(\Pr\left[I_{k,k+1}=1\right]\right)^{\frac{1}{N-1}} \notag
\end{eqnarray}
where (c) follows from $I_{k,k+1} \in \{0,1\}$, (d) follow from the fact that, conditioned on $\omega$, $I_{k,k+1}$ depends only on $t_k$ and $t_{k+1}$ and (e) is a definition of the probability, marginalized over the choices of $\omega$.  
\end{proof}

\section{Upper Bound on the Pairwise Probability of Correct Causal Ordering}
\label{sec:UpperBoundPairwise}

Here consider a more general setting to examine the impact of the TWI size on the pairwise probability of correct causal ordering. In this setup we have that the event ${\cal V}_c$ is equivalent to the event $I_{1,2}=0$. 
Assuming that the transmission times $T_1/T_2$ have probability density functions $p(t_1)/p(t_2)$, then the probability of causality violation under a given TWI size $W$ is calculated as follows. For given values $T_1=t_1$ and $T_2=t_2$ the probability of causality violation is calculated as follows:
\begin{equation}
    \Pr[{\cal V}_c|t_1, t_2]=
        \left\{\begin{array}{cc}
                0 & \text{if }t_1 \leq \tau+t_2 \\
                \frac{t_1-\tau-t_2}{W} & \text{if } \tau+t_2 < t_1 \leq \tau+t_2+W \\
                1 & \text{if }t_1 > \tau+t_2+W
        \end{array}\right.
\end{equation}
Using this, we can write: 
\begin{eqnarray}
    \Pr[{\cal V}_c]&=&\int_{0}^{\infty}p(t_2)\mathrm{d}t_2 \int_{\tau+t_2}^{\tau+t_2+W} \frac{t_1-\tau-t_2}{W} p(t_1)\mathrm{d}t_1 \notag \\
    &+& \int_{0}^{\infty}p(t_2)\mathrm{d}t_2 \int_{\tau+t_2+W}^{\infty} p(t_1)\mathrm{d}t_1 \notag \\
    &\stackrel{(a)}{\geq}& \int_{0}^{\infty}p(t_2)\mathrm{d}t_2 \int_{\tau+t_2+W}^{\infty} p(t_1)\mathrm{d}t_1
\end{eqnarray}
where (a) follows that the first double integral, containing $\frac{t_1-\tau-t_2}{W}$, is nonnegative. Assume that $T_1$ is distributed with an exponential tail, that is, there exists $\lambda$ such that 
\[\Pr[T_1>\tau+t_2+W]=e^{-\lambda (\tau+t_2+W)}\]
Then:
\begin{eqnarray}
    \Pr[{\cal V}_c] &\geq& \int_{0}^{\infty} e^{-\lambda (\tau+t_2+W)} p(t_2) \mathrm{d}t_2 \notag \\ 
    &=& e^{-\lambda (\tau+W)} \int_{0}^{\infty}e^{-\lambda t_2} p(t_2) \mathrm{d}t_2 \notag \\ 
    &=& e^{-\lambda (\tau+W)} E[e^{-\lambda t_2}]
\end{eqnarray}
where $E[e^{-\lambda t_2}]$ does not depend on $W$. We see that the impact of TWI is that it exponentially decreases the lower bound on the pairwise probability of causality violation. If we take $L=\frac{W}{\tau}$ as the event throughput loss, then the lower bound becomes proportional to $e^{-\lambda \tau(1+L)}$, which can be interpreted as follows: the linear event throughput loss causes an exponential decrease of the probability of causality violation.


\bibliographystyle{IEEEtran}
\bibliography{References}

\end{document}